# Compatibility of the selective area growth of GaN nanowires on AlN-buffered Si substrates with the operation of light emitting diodes

M Musolino, A Tahraoui, S Fernández-Garrido, O Brandt, A Trampert, L Geelhaar and H Riechert

Paul-Drude-Institut für Festkörperelektronik, Hausvogteiplatz 5–7, 10117 Berlin, Germany

E-mail: musolino@pdi-berlin.de

**Abstract**

AlN layers with thicknesses between 2 and 14 nm were grown on Si(111) substrates by molecular beam epitaxy. The effect of the AlN layer thickness on the morphology and nucleation time of spontaneously formed GaN nanowires (NWs) was investigated by scanning electron microscopy and line-of-sight quadrupole mass spectrometry, respectively. We observed that the alignment of the NWs grown on these layers improves with increasing layer thickness while their nucleation time decreases. Our results show that 4 nm is the smallest thickness of the AlN layer that allows the growth of well-aligned NWs with short nucleation time. Such an AlN buffer layer was successfully employed, together with a patterned $SiO_x$ mask, for the selective-area growth (SAG) of vertical GaN NWs. In addition, we fabricated light-emitting diodes (LEDs) from NW ensembles that were grown by means of self-organization phenomena on bare and on AlN-buffered Si substrates. A careful characterization of the optoelectronic properties of the two devices showed that the performance of NW-LEDs on bare and AlN-buffered Si is similar. Electrical conduction across the AlN buffer is facilitated by a high number of grain boundaries that were revealed by transmission electron microscopy. These results demonstrate that grainy AlN buffer layers on Si are compatible both with the SAG of GaN NWs and LED operation. Therefore, this study is a first step towards the fabrication of LEDs on Si substrates based on homogeneous NW ensembles.

Submitted to: *Nanotechnology*

## 1. Introduction

Light-emitting diodes (LEDs) based on III-N nanowires (NWs) are pursued for a number of reasons as an attractive alternative to their conventional planar counterparts [1, 2, 3]. NWs can be grown in high crystalline quality on dissimilar substrates, and particularly important are Si substrates due to their wide availability at low price and in large size. The reason for this advantage of NWs is that the NW geometry does not favour the vertical propagation of extended defects like dislocations [4]. Thus, this approach might pave the way for the monolithic integration of group-III-N- and Si-based technologies. Furthermore, the NW geometry allows the elastic relaxation of strain at the free sidewalls [5, 6], hence permitting the growth of (In,Ga)N/GaN heterostructures along the NW axis with higher In content than in the planar case. Such NW heterostructures are able to emit light in the entire visible spectral range from blue to red, and even in the infrared [7, 8, 9, 10]. Therefore, LEDs based on III-N NWs could enable solid state lighting without the need for phosphors [11, 12]. Several groups have already demonstrated the possibility to fabricate LEDs based on (In,Ga)N/GaN NWs grown on Si(111) with the help of self-assembly processes [2, 11, 13, 14, 15, 16, 17, 18, 19, 20]. However, the NW ensembles grown in such a way inevitably exhibit fluctuations in length and diameter of the NWs, that in turn cause differences in In incorporation. As a consequence, LEDs fabricated from such NW ensembles suffer from multicolour emission [11, 21] and strong inhomogeneities in the current path [18].

More homogeneous properties might be achieved by controlling the diameter and spacing of the NWs by selective-area growth (SAG) [7]. Another benefit of this approach is that nano-LEDs emitting at different colours may be fabricated on the same substrate in a single growth run [7, 22]. The SAG of III-N nanowires has been achieved by a few groups, both by plasma-assisted molecular beam epitaxy (PA-MBE) and by metal-organic vapour phase epitaxy (MOVPE), on GaN templates grown on sapphire [23, 24, 25, 26, 27]. Recently, the approach involving PA-MBE was extended to GaN buffers on Si [10]. Prior to this development, the SAG of GaN nanowires on Si substrates was demonstrated in an alternative way employing AlN buffers [28, 29, 30]. For a comparison of these two approaches on Si, we recall that when planar III-N layers are grown on Si substrates, the first step is typically the growth of an AlN buffer layer, because the mismatch in thermal expansion coefficient is much smaller than for GaN and because interfacial reactions can be suppressed more efficiently (during the direct growth of GaN on Si, the formation of $Si_xN_y$ and/or the melt back etching of Si by Ga are likely) [31]. At the same time, the large band gap of AlN presents a drawback for the overall series resistance of any electronic device fabricated on such structures. For the spontaneous growth of GaN NWs on Si by MOVPE, AlN buffers have been employed to induce NW nucleation. LEDs based on such NWs exhibited high threshold voltages exceeding 10V, but the origin of this high value was not analyzed further [17]. For either of the approaches to the SAG of III-N NWs on Si substrates it is, therefore, at present not clear whether functional LEDs can be obtained.

Here, we demonstrate the suitability of the approach involving AlN buffers. In order to minimize the resistance of the AlN buffer on the current flow through the LED, we reduce the thickness of this layer while maintaining its effectiveness for SAG. Then, we fabricate LEDs based on spontaneously formed NWs grown on such an AlN buffer layer to demonstrate that electrical conduction is possible.

## 2. Sample growth

The samples presented in this work were grown in MBE systems equipped with radio frequency plasma sources for the active N species and solid-source effusion cells for the metals and the dopants (*i. e.*, Ga, Al, In, Si and Mg). The reagent fluxes were calibrated in equivalent growth rate units for III-N planar layer (nm/min) as explained in Ref. [32]. The substrate temperatures reported in this work were acquired by an optical pyrometer if higher than 400°C and by a thermocouple in the case of lower values. The n-type Si(111) substrates were loaded into the growth chamber as-received, without any chemical treatment. In order to remove the native oxide, 100 monolayers (ML) of Ga were deposited at 550°C just before growth and subsequently desorbed by increasing the substrate temperature up to 800°C. This process was repeated until the reflection high-energy electron diffraction (RHEED) pattern characteristic for the 7×7 surface reconstruction was clearly observed. During growth, the group-III atoms desorbing from the substrate were monitored by line-of-sight quadrupole mass spectrometry (QMS); and the QMS response was also calibrated in equivalent growth rate units [33].

In order to obtain thin, uniform, and crystalline AlN buffer layers with a smooth surface directly on Si, we developed a three-step growth procedure that avoids the undesirable formation of $Si_xN_y$ and produces a closed and continuous AlN layer without any droplets. First, 1.2 ML of Al are deposited on the Si surface at a low temperature (50°C). The subsequent nitridation of this Al at the same temperature leads to the formation of a very thin AlN layer [34]. Then, AlN is grown at higher substrate temperature, 680°C, up to the desired total layer thickness by supplying simultaneously Al and N. During growth the III/V flux ratio is kept close to unity. This procedure allows the fabrication of smooth, uniform and close AlN layers on Si(111) even for thicknesses of only few nanometres.

## 3. Results and discussion

*3.1. Optimization of the thickness of the AlN buffer layer*

In order to minimize the series resistance of a NW-LED device on the AlN buffer, the thickness of this layer must be kept as thin as possible. At the same time, the GaN NWs grown on it must exhibit a good morphology and have to be well aligned perpendicular to the substrate surface, because it would otherwise not be possible to achieve the vertical contacting of the NW ensemble, and the controlled growth of an n-i-p junction and heterostructure that are well-stacked along the NW length would be seriously compromised.

In order to find the thinnest AlN layer that meets these requirements, a series of samples was prepared by growing GaN NWs on AlN layers of different thickness. In addition, to investigate the properties of the AlN surface as a function of the thickness, the same series of samples was reproduced without GaN NWs. The growth parameters of both the AlN layer and the GaN NWs were kept constant for all samples, which differ only in the growth time of the AlN layer. AlN layers with thicknesses between 2 and 14 nm were produced using the procedure described before, whereas the GaN NWs were grown at a substrate temperature of 805°C using fluxes equal to 4 and 10.5 nm/min for Ga and N, respectively (henceforth, represented by $\Phi_{Ga}$ and $\Phi_N$). The thickness of the AlN layers was determined by x-ray reflectivity measurements (data not shown here), while the crystalline structure and the morphology of its surface was analysed by RHEED and atomic force microscopy (AFM), respectively.

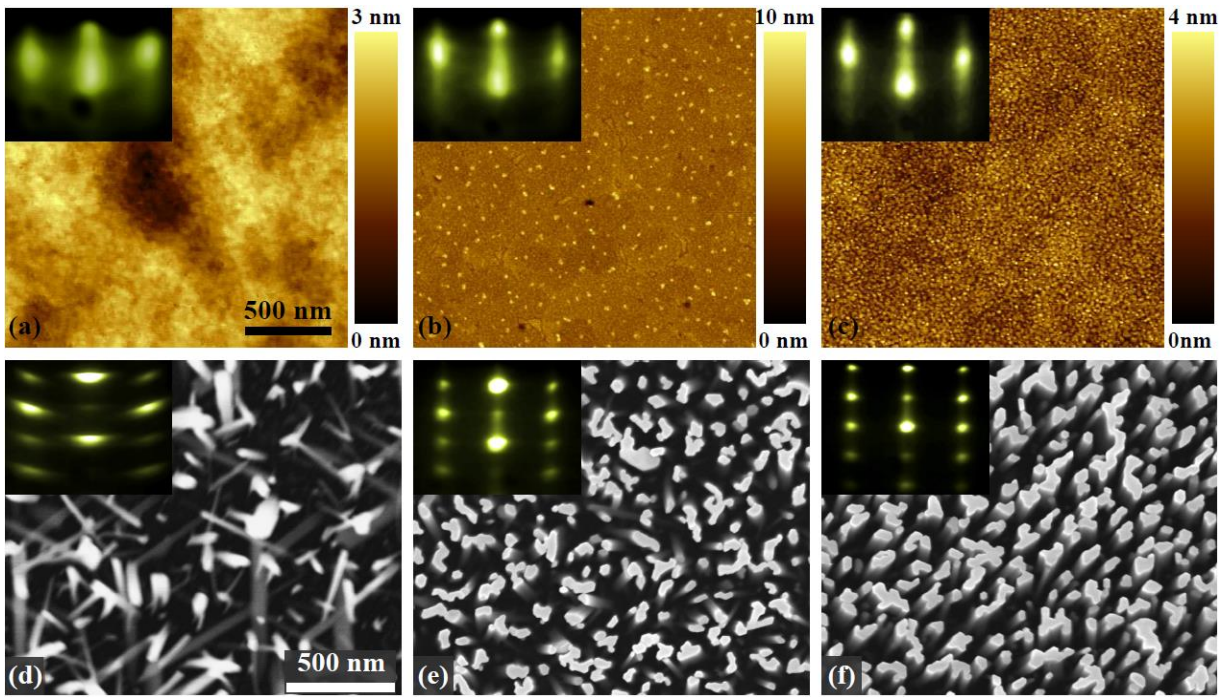

Figure 1. (a)–(c) AFM images of three AlN layers with different thicknesses: (a) 2 nm, (b) 4 nm, and (c) 8 nm. (d)–(f) Top-view SEM images of the GaN NWs grown on these AlN layers. The insets of the figures (a)–(c) and (d)–(f) depict the corresponding RHEED patterns acquired at the end of the growth along the [11$\bar{2}$0] azimuth, respectively.

Figures 1(a)–(c) show the AFM images of the surfaces of three AlN layers with different thicknesses (2, 4, and 8 nm), while the insets depict the corresponding RHEED patterns. All the three samples exhibit a very smooth surface with a root-mean-square

(RMS) value of the roughness of about 0.5 nm. Nevertheless, the surface morphology depends strongly on the thickness of the layer. In particular, the surface of the 2-nm-thick layer [figure 1(a)] is not completely flat, but modulated on the scale of a few hundreds of square nanometres by depressions, which in some points reach the substrate. The corresponding RHEED pattern shows broad and diffuse reflections, suggesting a poor crystalline quality. In contrast, the 4-nm-thick layer [figure 1(b)] is almost completely closed, and sporadic pits with average diameter of about 40 nm are visible. However, many small islands are observed. In this case, the RHEED pattern is brighter and better defined. Finally, the 8-nm-thick layer [figure 1(c)] shows a completely closed surface characterized by a very dense distribution of small islands. The corresponding RHEED pattern shown in the inset is more streaky than the ones related with the thinner layers, suggesting a better crystalline quality.

The alignment of the NWs with respect to the substrate was investigated *in situ* by RHEED and *ex situ* by scanning electron microscopy (SEM). Figures 1(d)–(f) show the SEM top view images of the GaN NWs grown on the three AlN layers described before, while the insets depict the corresponding RHEED patterns. As seen in figure 1(d), the NWs grown on the 2-nm-thick AlN layer are completely misaligned; and this result is also revealed by the RHEED pattern composed of reflections of oval shape arranged on circles, which indicates diffraction from crystallites oriented in different directions. Obviously, this AlN layer is not suitable for our purpose. The alignment of the NWs improved drastically when the thickness of the AlN layer was increased even by only 1 nm, and further improvements were obtained by increasing the thickness to 4 nm, as visible in figure 1(e). In this case, the RHEED pattern exhibits reflections of circular shape arranged on a rectangular grid and the SEM top view image reveals that the NWs are well-oriented perpendicular to the substrate surface, with only a small amount of inclined NWs. The shape and arrangement of these NWs is suitable for the realization of NW-based LEDs. With further increasing thickness of the AlN layer, the alignment of NWs improved more and more, as is visible in particular from the RHEED pattern with sharp reflections in figure 1(f) for the NWs grown on a 8-nm-thick AlN layer. The improved alignment of the GaN NWs with increasing thickness of the AlN layer is also confirmed by symmetric x-ray diffraction ω-scans, in which the full width at half maximum of the GaN(0002) reflection decreases for increasing thickness of the AlN layers (data not shown here).

Next, we investigate the suitability of these very thin AlN layers for SAG. Our approach to SAG [30, 35] is based on the observation that the spontaneous growth of GaN nanowires is preceded by a nucleation time during which there is no appreciable incorporation of Ga, indicating that nucleation of GaN does not occur [36, 37]. For SAG we exploit the fact that the nucleation time depends sensitively on the substrate material. In particular, for appropriate growth conditions the nucleation time can be extended to several hours on bare Si and on $SiO_x$, while it is much shorter on AlN. Good growth selectivity is achieved if the difference between the nucleation time of the GaN NWs on the mask ($SiO_x$ in our case) is much longer than on the underlying AlN buffer layer. Hence, the shorter the nucleation time on the AlN layer, the better is the selectivity. In order to determine which of the thin AlN layers considered here is suited for SAG, the same series of samples described above was used to study the nucleation time of the GaN NWs. To this end, we monitored *in situ* the flux of Ga desorbing from the substrate ($\Phi_{QMS}$) by QMS [35].

In figure 2 the QMS data that were acquired during the growth of GaN NWs on bare Si (grey curve) and on AlN layers with different thicknesses are plotted: the red, blue, and green profiles correspond to layer thicknesses of 3, 4, and 8 nm, respectively. Taking a careful look at the curve acquired during the growth of NWs on bare Si, three main regions are noticeable: i) an initial stage where $\Phi_{QMS}$ is nearly constant, associated with the incubation of the NWs; ii) a second stage where $\Phi_{QMS}$ decreases because of the formation of GaN NWs; iii) a final stage where $\Phi_{QMS}$ reaches steady-state conditions, related to the continuous elongation of the NWs. A similar behaviour has already been reported in previous studies carried out on Si substrates [38, 39, 40].

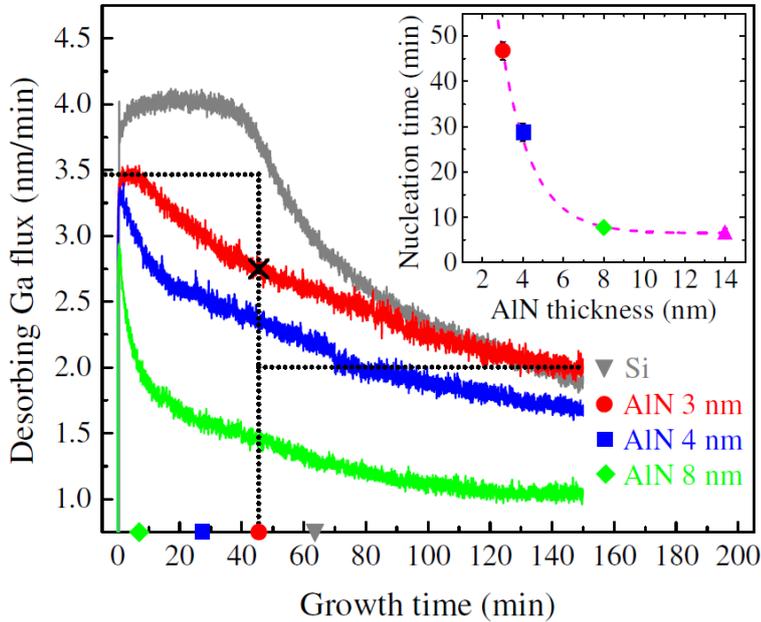

Figure 2. Desorbing Ga flux measured by QMS as a function of growth time for different starting surfaces: Si (grey profile) and AlN buffer layers that are 3 nm (red profile), 4 nm (blue profile), and 8 nm thick (green profile). The nucleation times in the inset were obtained from the QMS profiles as schematically sketched by the black dotted lines in the case of the red profile. The dashed line in the inset is a guide to the eye.

During the growth of GaN NWs on AlN layers, we observe comparable features. However, in these cases the incubation stages are much shorter than the one observed on Si, reflecting the fact that the nucleation of GaN NWs takes place faster on AlN than on Si. Moreover, this first stage becomes shorter and shorter with increasing AlN thickness. For instance, the incubation stage ends after only few tens of seconds in the case of the 8-nm-thick layer. Interestingly, immediately after the start of the deposition the samples on AlN show a lower desorption of Ga, although all the samples were grown under the same conditions. This observation might be explained by a higher sticking coefficient of the Ga atoms on the AlN surface due to its higher roughness which might offer nucleation centres for the formation of GaN islands.

Using the QMS data, we define the average nucleation time of the NWs as the time from the beginning of the deposition to the moment when the desorbed Ga flux decreases by half the difference between the values that it has during the incubation and the elongation stages, respectively [40]. This method is sketched by the dotted line in figure 2 for the QMS profile in red. Applying the same procedure also to the QMS profiles of the other samples, we extracted the data plotted in the inset of figure 2, that shows the nucleation time of the GaN NWs as a function of the thickness of the AlN layer. It is clear that the nucleation time strongly depends on the AlN thickness. In particular, the nucleation time decreases very fast with increasing thickness from 3 to 8 nm, whereas a further increase in AlN thickness does not lead to further significant changes in the nucleation time. In fact, the difference between the nucleation time on the 8 and 14-nm-thick AlN layers is only 1 s. However, in all the considered cases the nucleation time of the NWs grown on AlN is shorter than the one observed for NWs on bare Si which was as long as 62 min. The decrease of the nucleation time with increasing thickness of the AlN layers might be correlated with the higher density of surface irregularities observed in figure 1(a)–(c) for thicker AlN layers.

In conclusion of this subsection, the thinnest AlN layer that leads to the well-aligned growth of GaN NWs characterized by a relative short nucleation time is the one with a thickness of 4 nm. All further experiments were carried out with such AlN layers.

*3.2. Microstructure of the AlN buffer layer*

Since the AlN buffer layer is expected to influence the current flow through the envisioned NW-LED, it is important to investigate its microstructure. To this end, we analysed the interface region between the base of the NWs, the AlN buffer and the Si substrate by transmission electron microscopy (TEM). The high-resolution TEM (HRTEM) image in figure 3(a) shows a representative GaN NW grown on the 4-nm-thick AlN layer. Besides confirming the nominal thickness of the buffer layer, this image shows a fairly smooth interface between the Si substrate and the AlN, free of amorphous $Si_xN_y$. In addition, it becomes clear that the AlN layer is not a single crystal; on the contrary, it is composed of several grains with different crystallographic orientations, as highlighted by the yellow dashed lines in figure 3(a). A more detailed analysis of the grains reveals that the majority of them exhibits the hexagonal phase [see figure 3(b)], with some tilted and twisted domains [see figure 3(c)]. We found also a few inclusions of the cubic phase [see figure 3(d)]; cubic inclusions are likely to form during the growth of thin AlN layers at low temperature (lower than 700°C) [41, 42, 43]. Underneath the base of a NW, even if the NW is as thin as the one in figure 3(a) with a diameter of about 30 nm, typically several grains are visible. At this point we would like to emphasize that in our case the formation of grain boundaries in the AlN layer is actually desirable. In fact, such defects are known to facilitate charge transfer across insulating layers [44]. Thus, for the envisioned NW-LED it is favourable that the AlN buffer enables the vertical growth of NWs and exhibits at the same time a high density of conductive paths for electrons.

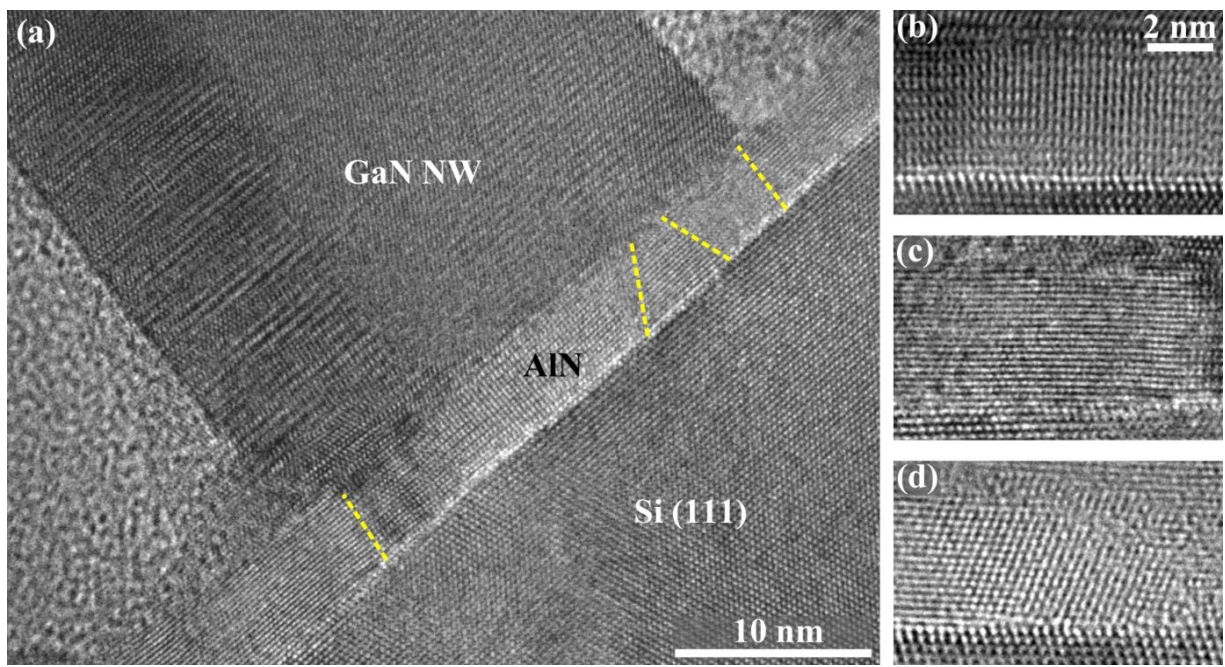

Figure 3. (a) HR-TEM image of a GaN NW grown on the 4-nm-thick AlN layer. The yellow dashed lines highlight the boundaries of different grains in the AlN layer. (b)–(d) Magnification of three different grains with hexagonal (b), twisted hexagonal (c) and cubic (d) phase.

*3.3. SAG of GaN NWs*

In previous SAG studies, AlN buffers with a thickness of about 10 nm were employed [30, 45]. As the next step we have therefore to prove that also the 4-nm-thick AlN layer on Si enables the SAG of GaN NWs with good selectivity and NW morphology, in particular in view of the grainy structure discussed in the previous subsection. For this aim, a 10-nm-thick $SiO_2$ layer was deposited by magnetron sputtering on AlN/Si samples. Then, these samples were coated with a lithographic resist and patterned by soft UV nanoimprint lithography. The resulting pattern in the resist was then transferred into the $SiO_2$ mask by reactive-ion etching, resulting in regular arrays of holes in the mask to the underlying AlN buffer. The patterned substrates were finally cleaned with solvents in an ultrasonic bath to remove the residual resist just before the insertion into the MBE chamber.

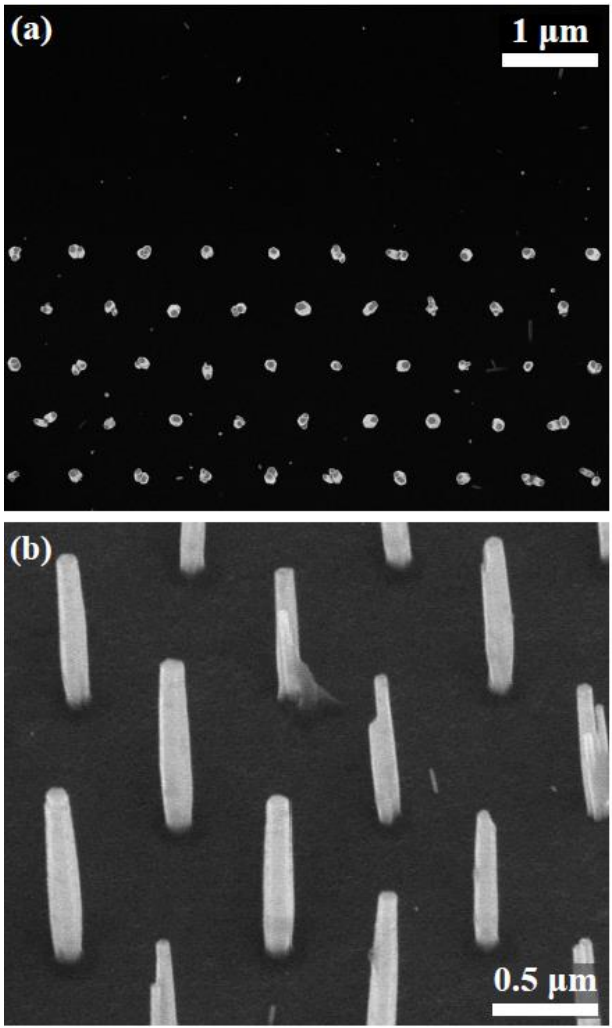

Figure 4. Top (a) and bird's-eye view (b) SEM images of selectively grown GaN NWs with different arrangements. The diameter of the holes is 60 nm while the distance between their centres is 0.7 and 1 µm for (a) and (b), respectively.

The growth conditions used for the SAG of GaN NWs were comparable with the ones used for self-induced growth: $\Phi_{Ga}$=3 nm/min, $\Phi_N$=10 nm/min and a substrate temperature of 813°C. Further details about the optimization of the growth parameters for the SAG of GaN NWs using a $SiO_2$ mask on an AlN buffer layer can be found elsewhere [30]. The sample presented here was grown for 3 hours. After growth the sample was analysed by SEM; figure 4 presents a top view (a) and a bird's-eye view (b) of different arrays of SAG

NWs. Figure 4(a) shows that the selectivity is good even outside the patterned area; only very few and thin NWs spontaneously formed on the mask. In between the regular array of NWs the parasitic nucleation of GaN on the mask is suppressed even better, also when the distance between the NWs is as large as 1 µm, as shown in figure 4(b). In almost all of the holes with diameters in the range of 40–70 nm, one single NW is formed. In larger holes, multiple nucleation took place leading to the formation of separated NWs that are expected to coalesce into single NWs with continued growth. In this respect, it is worth noting that only in thin NWs, for which the elastic relaxation of strain at the free sidewall surfaces is efficient, a high crystal quality and high In content of axial (In,Ga)N/GaN heterostructure is possible [5, 6]. Almost all of the NWs are oriented perpendicular to the substrate, have a hexagonal C-plane top facet, and exhibit rather homogeneous heights and diameters. Therefore, the SAG of GaN NWs is possible on Si substrates even with an AlN buffer layer that is only 4 nm thick.

*3.4. NW-LED on AlN-buffered Si*

As the last step of this study, we have to verify whether the 4-nm-thick AlN layer is thin enough to allow an electric current to flow and therefore makes the operation of an LED possible. To this end, we fabricated LEDs based on NWs spontaneously formed on the 4-nm-thick AlN buffer layer on a Si substrate. Of course, the final goal will be to realize an LED structure with SAG and process the resulting homogeneous NW ensemble into a device. However, towards this aim several optimization steps would still be needed. For example, SAG has to be combined with doping as well as the formation of heterostructures, and the processing procedure has to be adjusted to the new ensemble geometry. Such efforts go beyond the scope of the present study that focuses on the compatibility of the AlN buffer with both SAG and LED operation. In this subsection we will also investigate the effect of the AlN layer on the optoelectronic characteristics of the NW-LEDs. To this end, the performance of LEDs grown on the 4-nm-thick AlN layer and directly on bare Si were compared.

The LEDs based on spontaneously formed NWs were produced following the processing scheme described in Ref. [18] and [20]. The active regions of the LEDs consist of four (In,Ga)N quantum wells (QWs) with an average In content of about 20%, separated by three GaN barriers. They are embedded between two doped GaN segments designed such that an n-i-p diode doping profile is created. We point out that the NWs in the samples grown for this study on thin AlN and bare Si have comparable morphology and density. To fabricate the final devices, the NW ensembles were planarized by spin coating, and then a transparent 120-nm-thick indium-tin-oxide p-type contact was sputtered on the top of the samples. Finally, the n-type contacts were created by depositing Al(100 nm)/Au(50 nm) on the back side of the n-Si substrate.

As the first step to compare the performance of NW-LEDs grown on AlN-buffered and bare Si, we acquired electroluminescence (EL) maps of the samples fabricated on these two types of substrates. Figures 5(a) and (b) show maps of the resulting EL acquired through an optical microscope on devices with an area of 0.2 mm$^2$ and an applied forward bias of 6V. Both the NW-LEDs grown on AlN and on bare Si exhibit spotty emission patterns, with a comparable density of emitting spots. The emission of EL from the sample grown on the 4-nm-thick AlN layer clearly proves that an electric current can flow across the barrier formed by this layer. Figure 5(c) shows the density of EL spots extracted from the EL maps as a function of the applied forward bias. The turn-on of the first individual NW-LEDs occurs in both the samples at similar bias (about 2 V), then the number density of emitting NWs rises faster in the sample grown on bare Si than in the one grown on thin

AlN. Beyond 6 V the density of EL spots saturates at about $2\times10^8$ cm$^{-2}$ regardless of the presence of the thin AlN layer.

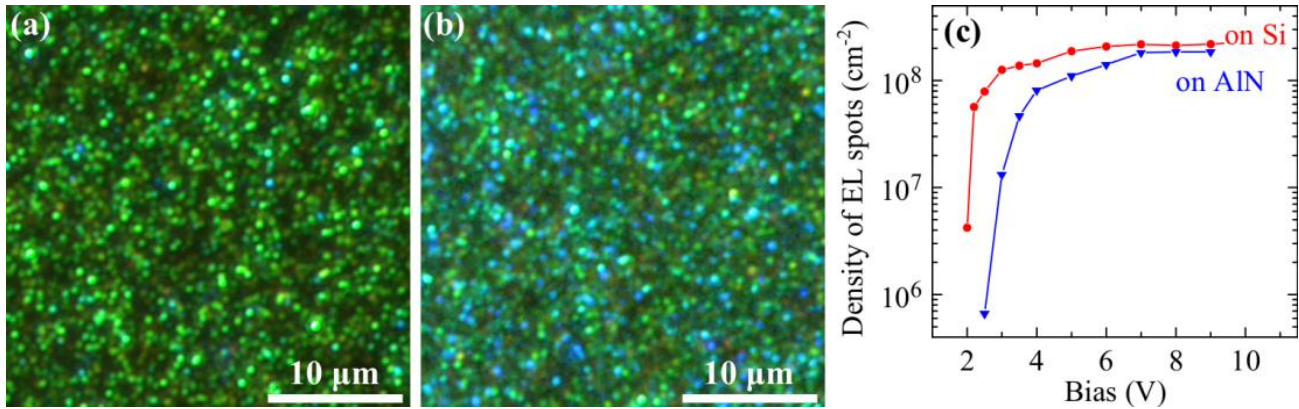

Figure 5. EL maps of the NW-LEDs grown on (a) 4-nm-thick AlN, and (b) directly on Si under a forward bias of 6 V acquired through an optical microscope. (c) Dependence of the number density of EL spots on the forward bias for the two different types of NW-LED.

Next, we compare the optoelectronic characteristics of the NW-LEDs grown on the two different substrates. Figures 6(a) and (b) depict the room temperature EL spectra acquired at different injected currents for the samples grown on AlN-buffered and bare Si, respectively. In all spectra, broad peaks centred in the green spectral range around 560 nm are observed. There are slight differences in the width and in the shape of the EL peaks between the two samples. These differences are not surprising for two different growth runs involving self-organization phenomena. In any case, the evolution of the EL spectra with the current is comparable. Figure 6(c) shows the EL intensity integrated over the entire spectral range as a function of the injected current. Surprisingly, the EL intensity emitted by the devices grown on thin AlN and on bare Si is comparable, and for currents higher than 20mA the NW-LED grown on AlN is even brighter. This result is also confirmed by figure 6(d), where the integrated EL intensity is plotted as a function of the power consumption, *i. e.*, the product of injected current and applied voltage. It is clearly visible that the devices grown on thin AlN and on bare Si emit the same EL intensity in the entire range of considered power.

In order to further understand the effect of the 4-nm-thick AlN layer on the flow of an electric current, we next present the analysis of the current-voltage (I-V) characteristics of the samples fabricated on the two types of substrates. Figure 6(e) depicts the I-V characteristics of the two studied samples on a linear scale. The sample grown on bare Si exhibits a turn-on voltage that is about 2 V lower than the one observed for the device grown on the thin AlN layer. The lower turn-on voltage of the sample grown on bare Si might be due to several factors. Firstly, the number of individual NW-LEDs in operation for bias lower than 6 V differs, and hence the total current conducted is lower in the device grown on the AlN layer than in the one grown on bare Si, as clearly visible in figure 5(c). Therefore, for small biases, the device on bare Si conducts an higher total current. Secondly, an early turn-on voltage might be caused by a high leakage current. Indeed, figure 6(f) reveals that for reverse bias the sample grown on bare Si has a leakage current about one order of magnitude higher than the one found in the sample with the AlN layer. Assuming that a similar leakage current occurs also for positive biases, for equal injected current the device grown on bare Si would emit EL less efficiently. These arguments might explain the very similar optoelectronic performance observed in the two devices, *i. e.*, comparable efficiencies at a given current and power consumption, despite the higher turn-on voltage of the sample grown on AlN layer.

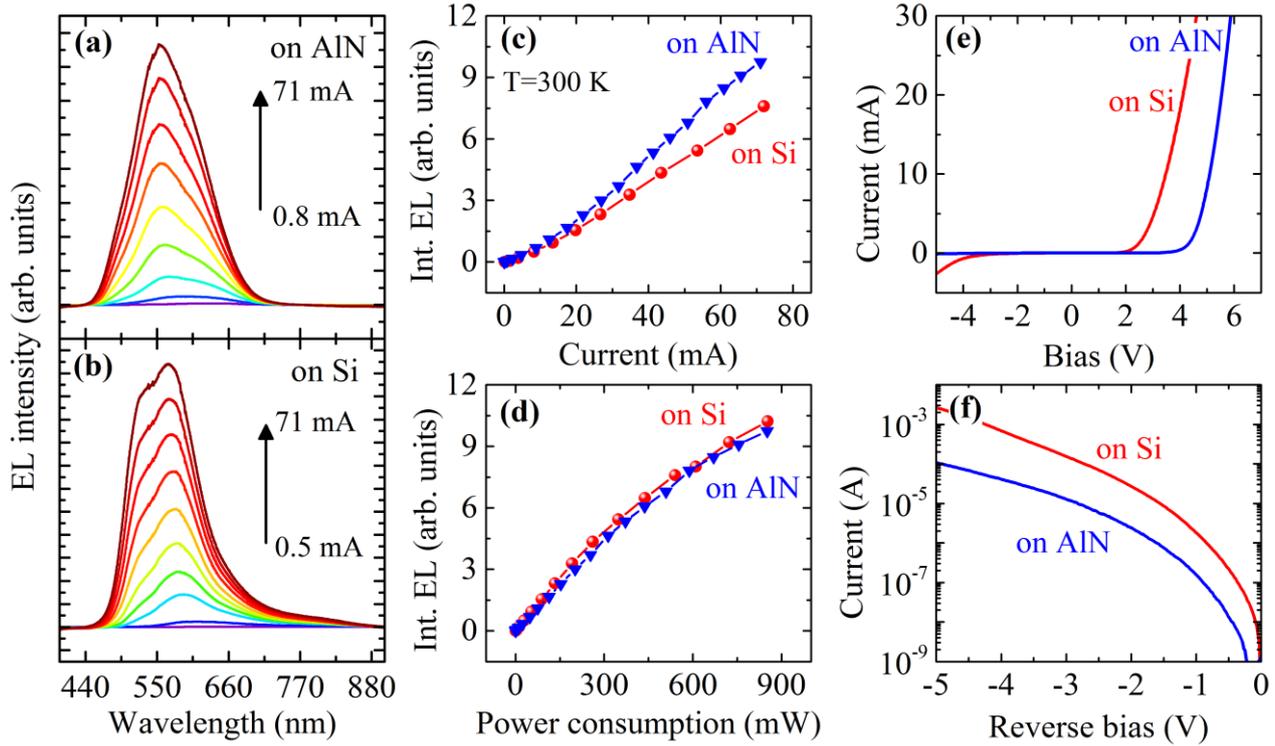

Figure 6. Optoelectronic characteristics of the NW-LEDs with area of 0.2 mm$^2$ fabricated on the two types of substrates. Room-temperature EL spectra acquired at different injected currents for the NW-LEDs grown on (a) 4-nm-thick AlN and (b) bare Si. Integrated EL intensity plotted versus (c) the injected current and (d) the power consumption. (e) Current-voltage characteristics of the two studied samples on a linear scale. (f) Same curves as in (e) plotted on a logarithmic scale versus the reverse bias.

The reasonably good transport properties observed in the NW-LED grown on the 4-nm-thick AlN layer would not be possible if the AlN was both closed and single crystalline. Two different structural features can provide conductive paths for the current driving the EL: on the one hand, the small pits in the AlN layer observed by AFM [see figure 1(b)], and on the other hand, the grain boundaries found by TEM [see figure 3(a)]. However, the number density of the pits is about seven times lower than the one of the EL spots, and the diameter of the pits is so small that only a single NW could be contacted through one pit. Therefore, the grainy structure of the AlN buffer is most likely the crucial factor for understanding the similar EL characteristics of the two types of NW-LEDs.

## 4. Summary

We have developed a three-step growth procedure for the production of thin and closed AlN layers without any droplets. On this basis, we have demonstrated that an AlN buffer layer that is only 4 nm thick both enables the SAG of GaN NWs on Si substrates and permits the operation of NW-LEDs. A careful characterization of the optoelectronic properties of comparable devices fabricated either on AlN-buffered or on bare Si has shown that the performance of NW-LEDs on bare and AlN-buffered Si is similar. The flow of electric current across the AlN buffer is facilitated by the grainy structure of this layer. This study is a first step towards the fabrication of NW-LEDs on Si substrates with homogeneous properties of the NW ensemble.

**Acknowledgments**

The authors gratefully acknowledge F. Limbach and T. Gotschke for scientific advice, B. Drescher and W. Anders for their help with the fabrication of the devices, C. Herrmann and H.-P. Schönherr for the maintenance of the MBE systems, and N. Koo as well as J. Kim from AMO GmbH for providing the patterned substrates.